\documentclass{ws-procs961x669}
\usepackage[mathscr]{euscript}
\usepackage{array,multirow,makecell}
\usepackage[breaklinks, colorlinks, citecolor=blue]{hyperref}
\usepackage{url,doi}
\usepackage{graphicx}
\usepackage{bm}
\usepackage[numbers]{natbib}

\begin{document}

\wstoc{Using \textit{Planck} maps for a systematic search of ultra-bright high-redshift strongly lensed galaxies}
{Matteo Bonato; Leonardo Trobbiani; Ivano Baronchelli; Gianfranco De Zotti; Mattia Negrello; Tiziana Trombetti; Carlo Burigana, Vincenzo Galluzzi and Erlis Ruli}

\title{Using \textit{Planck} maps for a systematic search of ultra-bright high-redshift strongly lensed galaxies}

\author{Matteo Bonato;$^{1,a}$
Leonardo Trobbiani;$^{2}$
Ivano Baronchelli;$^{1}$
Gianfranco De Zotti;$^{3}$
Mattia Negrello;$^{4}$
Tiziana Trombetti;$^{5}$
Carlo Burigana;$^{5,6}$
Vincenzo Galluzzi$^{5}$ and
Erlis Ruli$^{7}$}
\address{$^{1}$INAF - Istituto di Radioastronomia and Italian ALMA Regional Centre, Via Piero Gobetti 101, 40129 Bologna, Italy\\
$^{2}$Dipartimento di Fisica e Astronomia, Alma Mater Studiorum Universit\`a di Bologna, via Piero Gobetti 93/2, I-40129, Bologna, Italy\\
$^{3}$INAF - Osservatorio Astronomico di Padova, Vicolo dell’Osservatorio 5, I-35122 Padova, Italy\\
$^{4}$School of Physics and Astronomy, Cardiff University, The Parade, CF24 3AA, UK\\
$^{5}$INAF - Istituto di Radioastronomia, Via Gobetti 101, I-40129, Bologna, Italy\\
$^{6}$INFN, Sezione di Bologna, Via Irnerio 46, 40126, Bologna, Italy\\
$^{7}$Dipartimento di Scienze Statistiche, Universit\`a degli Studi di Padova, Via Cesare Battisti 241/243, I-35121 Padova, Italy\\
\vskip 0.1cm
$^{a}$matteo.bonato@inaf.it}

\begin{abstract}
This paper presents a novel approach to the use of \textit{Planck} telescope data for the systematic search of ultra-bright high-redshift strongly lensed galaxies. These galaxies provide crucial insights into the early universe, particularly during phases of intense star formation. The \textit{Planck} mission, despite its limited angular resolution, offers a unique opportunity to identify candidate strongly lensed galaxies over a wide area of the sky. This work outlines the methodology used to identify these rare objects, the challenges encountered, and the preliminary results obtained from follow-up observations with higher-resolution instruments.
\end{abstract}


\bodymatter

\section{Introduction}\label{sec:intro}

One of the most significant discoveries from the \textit{Planck} surveys was the unexpected identification — although predicted by  Ref.~\citenum{Negrello2007} — of ultra-bright, strongly lensed high-redshift sub-millimeter galaxies with extreme magnifications, $\mu$, ranging from 10 to 50 (see  Refs.~\citenum{Canameras2015} and \citenum{Harrington2016}). The \textit{Planck} telescope, renowned for its comprehensive mapping of the Cosmic Microwave Background (CMB), provides an additional exciting avenue for identifying these lensed galaxies. Although \textit{Planck} lacks the angular resolution to directly observe the detailed structure of lensed galaxies, its all-sky coverage enables the detection of these rare objects, which are subsequently confirmed through higher-resolution follow-up observations. These galaxies present a unique opportunity to explore the internal structure and kinematics of high-$z$ galaxies during their most active, dust-enshrouded star formation phases.

Understanding these details is crucial for deciphering the key processes that drive galaxy formation and early evolution. Current galaxy formation models propose a variety of physical mechanisms, such as mergers, interactions, cold flows from the intergalactic medium, and in situ processes (see reviews by  Refs.~\citenum{Silk2012} and \citenum{Somerville2015}). These models are often {characterized by} numerous adjustable parameters, enabling them to align with existing statistical data, including source counts and redshift distributions.

To directly {probe} the physical processes at work, it is essential to {observe} the interiors of high-$z$ star-forming galaxies. However, these galaxies are typically compact, with sizes ranging from 1 to 2 kpc (e.g.,  Ref.~\citenum{Fujimoto2018}), which correspond to angular sizes of 0.1 to 0.2 arcseconds at redshifts of $\sim$2 to 3. Resolving these small scales is challenging, even when using extremely powerful telescopes like the Atacama Large Millimeter/submillimeter Array (ALMA), the James Webb Space Telescope (JWST), and the Hubble Space Telescope (HST). When these galaxies are resolved, adequate signal-to-noise ratios per resolution element are usually achieved only for the brightest galaxies, which may not represent the general population.

Strong gravitational lensing offers a solution by enabling the study of high-$z$ galaxies in extraordinary detail, beyond the capabilities of current instrumentation. This is achieved through the magnification of the galaxy’s flux combined with the stretching of its images. Since lensing conserves surface brightness, the effective angular size is typically stretched by a factor of $\mu^{1/2}$.

ALMA observations, with a resolution of $0.1''$, achieved an impressive spatial resolution of approximately 60 pc, significantly smaller than the size of Galactic giant molecular clouds.  Ref.~\citenum{Canameras2017} also obtained CO spectroscopy, allowing them to measure the kinematics of the molecular gas with an uncertainty of $\sim$40 to 50$\, \rm km/s$. This level of spectral resolution enables the direct investigation of massive outflows driven by Active Galactic Nucleus (AGN) feedback at high $z$, which are predicted to have velocities of around $1000\, \rm km/s$.\cite{King2015}

Outflows are a key component in all major galaxy formation models, invoked to explain the inefficiency of star formation in galaxies (where only about 10\% of baryons form stars). However, observational evidence for these outflows at high $z$ has been scarce. Strong lensing enabled  Ref.~\citenum{Spilker2018} to detect a fast ($800\, \rm km/s$), massive molecular outflow in a galaxy at $z = 5.3$.

The all-sky, shallow \textit{Planck} surveys have identified the brightest high-$z$ strongly lensed sub-mm galaxies, making them ideal candidates for high spatial and spectral resolution follow-up studies. However, to draw reliable conclusions about galaxy formation physics, a statistically significant sample across a broad redshift range is necessary. To date, some tens of strongly lensed galaxies detected by \textit{Planck} have been identified, mostly through \textit{Herschel} DDT "Must-Do" Programmes (see  Ref.~\citenum{Canameras2015}) and by cross-matching \textit{Planck} catalogues with large-area sub-mm surveys.\cite{Berman2022} Given that these searches were conducted over quite limited sky areas (generally a few square degrees), it is likely that many more such objects remain undiscovered. A rough estimate of the expected number of \textit{Planck}-detected strongly lensed galaxies in the high-Galactic latitude ($|b|> 20^\circ$) sky can be obtained by noting that cross-matching \textit{Planck} with \textit{Herschel} surveys covering approximately 1000\,deg$^{2}$ yielded five such objects; extrapolating to the total area yields a total of $\sim 150$. Finding them would provide an excellent sample for investigating the structure of galaxies at and beyond the peak of cosmic star formation.

The structure of this paper is as follows. In Section \ref{sect:Planck}, we detail the methodology employed to select our strongly lensed galaxy candidates from \textit{Planck} maps. Section \ref{sect:challenges} presents the follow-up observations conducted on these candidates. Finally, we present our conclusions in Section \ref{sect:conclusions}.

\section{Methodology}\label{sect:Planck}

The all-sky maps generated by \textit{Planck} contain thousands of sub-millimeter sources, {a small fraction of which} are expected to be strongly lensed galaxies. Being dusty galaxies, the most suitable \textit{Planck} channels {to look for them} are at frequencies $\nu \geq$353 GHz. As predicted by  Ref.~\citenum{Negrello2007} and later confirmed through the analysis of H-ATLAS data\cite{Negrello2017}, at the bright flux density limits of \textit{Planck} surveys unlensed dusty galaxies are typically found at redshifts $z\leq 0.1$ (see also  Ref.~\citenum{Negrello2013}), while lensed galaxies are located at $z>1$. Consequently, the latter galaxies exhibit significantly colder (i.e. redder) sub-mm colors compared to local dusty galaxies.

%%%%%%%%
% FIGURE %
%%%%%%%%
\begin{figure}
\begin{center}
\includegraphics[trim=1.0cm 1.0cm 1.0cm 1.0cm, width=0.99\textwidth]{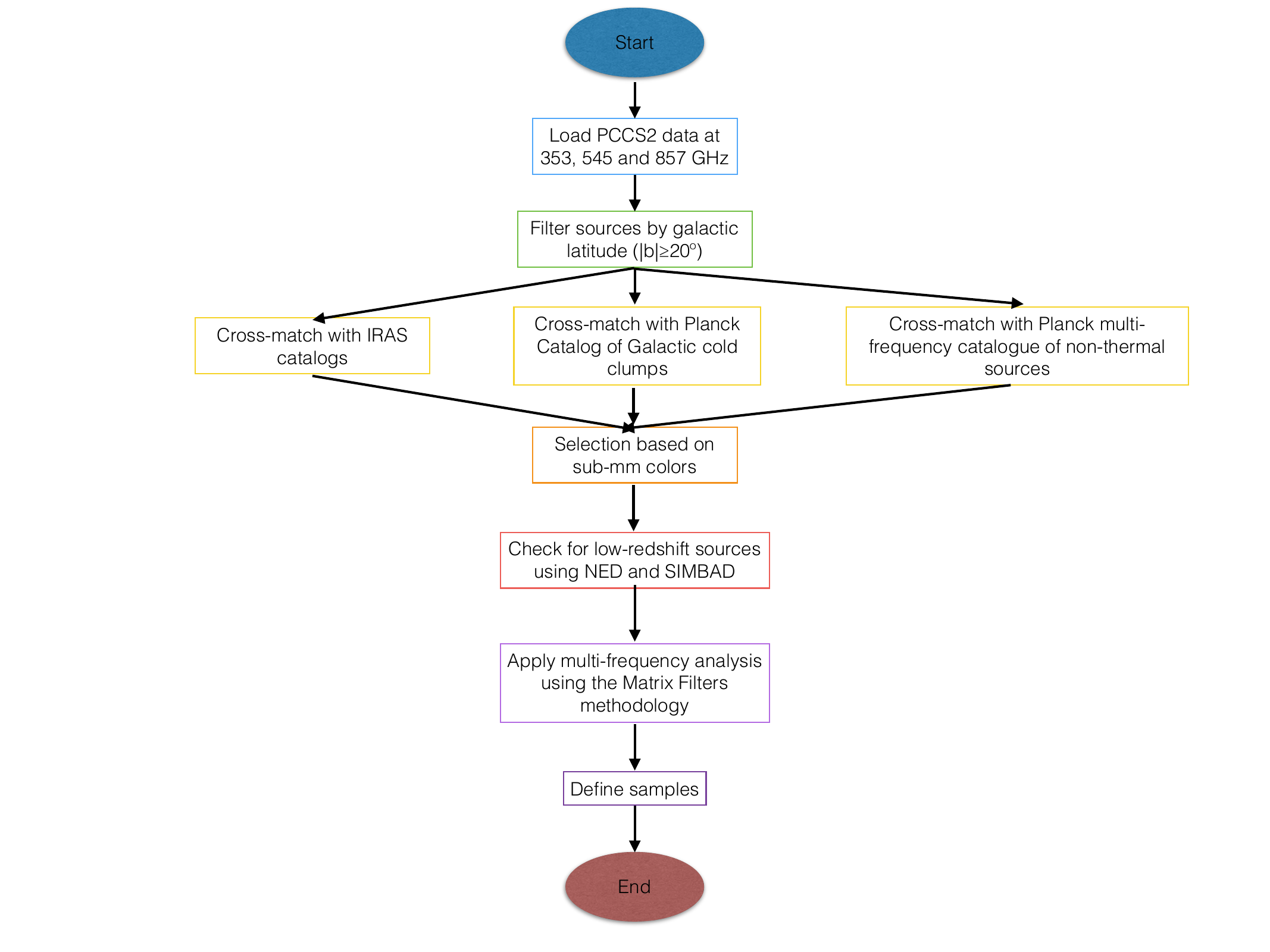}
\caption{Flowchart showing the main steps of our procedure for the definition of the sample of lensed galaxy candidates.}
 \label{fig:Flowchart}
  \end{center}
\end{figure}

{The procedure for the selection of the candidate lensed galaxies sample} from the \textit{Planck} maps {is} presented in  Ref.~\citenum{Trombetti2021}{. Here} we summarise the main steps{,} graphically represented in Fig.\,\ref{fig:Flowchart}.

We began by using the Second \textit{Planck} Catalogue of Compact Sources (PCCS2)\cite{PlanckCollaborationXXVI2016} at the three highest frequencies (353, 545, and 857 GHz), selecting only sources with Galactic latitudes $|b|>20\,$deg (to minimize contamination from Galactic emissions). Next, we performed an initial filtering of nearby galaxies by cross-matching our sample with the Infrared Astronomical Satellite (IRAS) Point Source and Faint Source catalogues, which primarily consist of low-redshift galaxies ($\sim$95\% of the total). Additionally, a cross-matching analysis with the \textit{Planck} Catalogue of Galactic cold clumps\cite{PlanckCollaborationXXVIII2016} enabled us to eliminate Galactic dusty clouds whose colors resemble those of high-redshift galaxies. We also eliminated radio sources by cross-matching with the \textit{Planck} multi-frequency catalogue of non-thermal sources (PCNT)\cite{PlanckCollaborationPCNT}.

Following this, we conducted a selection based on sub-mm colors, using the confirmed \textit{Planck} strongly lensed galaxies as templates. To further refine our sample, we checked for any low-redshift sources that might have escaped previous filters by using the NASA/IPAC Extragalactic Database (NED) or SIMBAD. For the remaining candidates, we applied a multi-frequency analysis using the "Matrix Filters" methodology (described in  Ref.~\citenum{PlanckCollaborationPCNT}). This approach allowed us to improve the signal-to-noise ratios (SNRs) compared to those of the PCCS2 and to estimate flux densities or set upper limits at \textit{Planck} frequencies where the PCCS2 provided no data.

%%%%%%%%
% FIGURE %
%%%%%%%%
\begin{figure}
\begin{center}
\includegraphics[trim=2.0cm 1.0cm 2.0cm 1.0cm, width=0.99\textwidth]{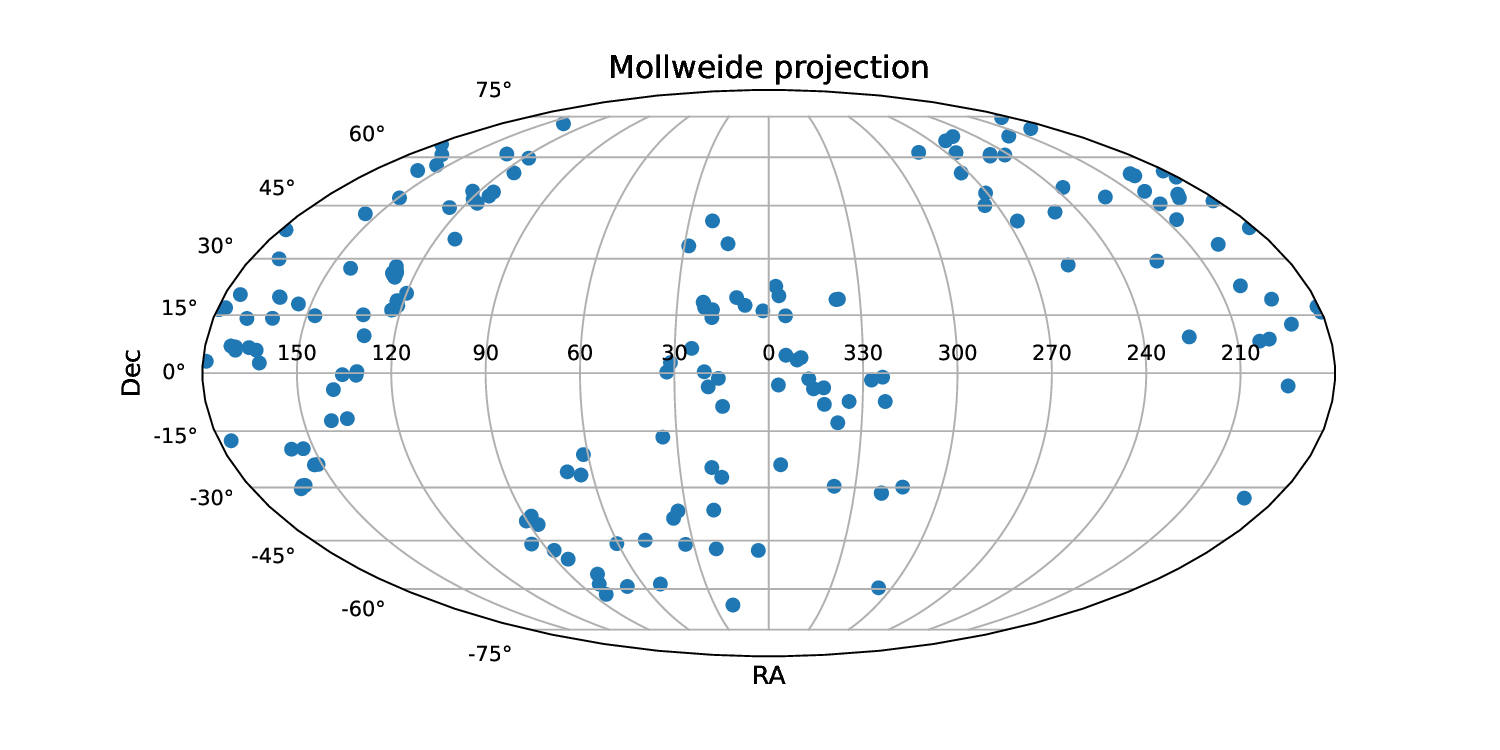}
\caption{Mollweide projection of the spatial distribution of our candidate strongly lensed galaxies selected from the 545\,GHz \textit{Planck} catalog.}
 \label{fig:Mollweide}
  \end{center}
\end{figure}

We have defined samples of lensed candidates selected at {353, 545, and 857} GHz, comprising {104, 177, and 97} distinct sources, respectively. An example of the distribution on the sky of our candidates selected at 545\,GHz is shown in Fig.\,\ref{fig:Mollweide}.

Alternatively, or as a complement to our approach, machine-learning techniques could be employed to select lensed candidates. This would involve extensive simulations, using tools like the \textit{Planck} Sky Model, to generate a customized training set.

\section{Challenges and Follow-Up Observations}\label{sect:challenges}

It is likely that more than 50\% of the sources in our samples are not strongly lensed galaxies, but rather a mixture of other objects with cold spectral energy distributions. These include high-redshift proto-clusters of dusty galaxies, Galactic cold clumps, cosmic infrared background (CIB) fluctuations, and Galactic cirrus.

{Identifying strongly lensed galaxies using \textit{Planck} data alone is nearly impossible. A major challenge arises from the low angular resolution of \textit{Planck}, resulting in significant positional uncertainties. These faint, low SNR sources have positional errors of at least one arcminute. To obtain the precise positions necessary for the highest-resolution telescope observations and detailed studies, we conducted follow-up observations with:} the Australia Telescope Compact Array (ATCA) at 5.5 GHz (project ID C3301, P.I.: M. Bonato); the second generation Neel-IRAM-KID-Array (NIKA 2), at 1 and 2 mm (project ID
212-19, P.I.: M. Bonato); the Submillimeter Common-User Bolometer Array 2 (SCUBA-2) at the James Clerk Maxwell Telescope (JCMT), at 850 $\mu$m (proposal ID: M19BP010, P.I.: M. Negrello). Four of the detected sources were observed spectroscopically with the Northern Extended Millimeter Array (NOEMA; Proposal ID: S20BQ, P.I.: M. Negrello). Each object of our sample has been observed by at least one of such facilities. These observations confirm the presence of sources and provide accurate positional determinations, necessary to apply for high-resolution {campaigns} with telescopes like ALMA, JWST, and the Karl G. Jansky Very Large Array (JVLA).

%%%%%%%%
% FIGURE %
%%%%%%%%
\begin{figure}
\begin{center}
\includegraphics[trim={0 0 0 0.65cm},clip, width=0.49\textwidth]{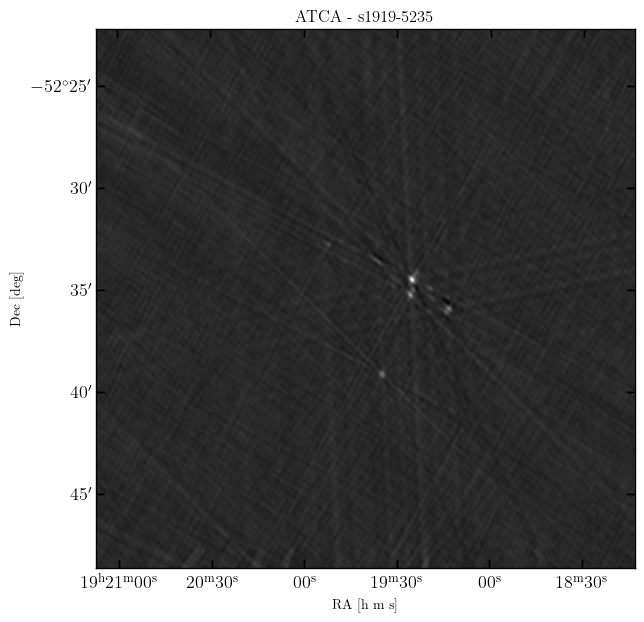}
\includegraphics[trim={0 0 0 0.6cm},clip, width=0.48\textwidth]{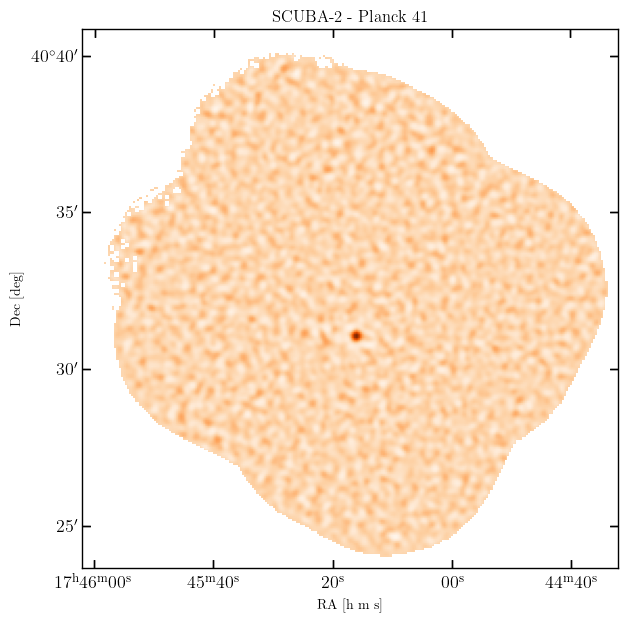}
\caption{{Examples of 5.5 GHz ATCA (\textit{left}) and 850 $\mu$m SCUBA-2 (\textit{right}) images from our follow-up observations of strongly lensed galaxy candidates. The ATCA image reveals multiple source detections at SNR$\geq$5, while the SCUBA-2 image shows only one at that significance level.}}
 \label{fig:images}
  \end{center}
\end{figure}

The analysis of these follow-up observations is ongoing. A couple of examples of ATCA and SCUBA-2 images, revealing the presence of SNR$\geq$5 source detections, are shown in Fig.\,\ref{fig:images}. Some preliminary results, which will be presented in Trobbiani et al. (in prep.), indicate the successful identification of new strongly lensed galaxies. To fully understand the nature of most of the detected sources, which could be lensed galaxies, proto-clusters, or other objects, higher resolution observations with ALMA, JWST, or JVLA are required. These findings will contribute to a more comprehensive understanding of the internal structure, star formation rates, and kinematics of high-redshift galaxies.

\section{Conclusions}\label{sect:conclusions}

The \textit{Planck} mission, while primarily designed for cosmological studies, has proven to be a valuable tool in the search for ultra-bright high-redshift strongly lensed galaxies. The systematic approach developed in this study, combined with follow-up observations using higher-resolution instruments, has the potential to significantly expand the sample of known lensed galaxies. This expanded sample will provide crucial data for studying galaxy formation and evolution during the peak of cosmic star formation activity, providing direct insight into physical processes that would otherwise remain inaccessible.

However, identifying candidate strongly lensed galaxies in \textit{Planck} catalogues is challenging because they are typically detected with low signal-to-noise ratios, with the exception of the few brightest sources. As a result, their photometric properties are significantly blurred, making it difficult to isolate them. To address this, we developed a method capable of increasing the number of identified \textit{Planck}-detected strongly lensed galaxies by a factor of approximately three to four, though with necessarily limited efficiency.
Our approach takes advantage of the fact that strongly lensed galaxies have colder sub-millimeter colours compared to nearby dusty galaxies, which constitute the vast majority of extragalactic sources detected by \textit{Planck}. Additionally, most nearby galaxies and radio sources can be excluded by cross-matching with the IRAS and PCNT catalogues, respectively.

We are analyzing higher-resolution follow-up photometric and spectroscopic observations of our candidates made with ATCA, NIKA 2, SCUBA-2 and NOEMA. These observations confirm the existence of the sources and provide precise positional data, enabling detailed high-resolution follow-up studies with telescopes such as ALMA, JVLA, and JWST.

Future work will focus on refining the selection process, exploring alternative methods such as machine learning, and conducting detailed studies of the confirmed lensed galaxies.

\section*{Acknowledgments}
We acknowledge support from INAF under the mini-grant ``A systematic search for ultra-bright high-$z$ strongly lensed galaxies in \textit{Planck} catalogues''.

%\section{References}

\bibliographystyle{ws-procs961x669}
%\bibliography{biblio.bib}

%\begin{verbatim}

%\end{verbatim}

\end{document}